\documentstyle[11pt]{article}

\begin{document}

\title{Higgs Pain? Take a Preon!
\footnote{Presented by S.F. at the Joint Meeting
of the Networks 'The Fundamental Structure
of Matter' and 'Tests of the Electroweak
Symmetry Breaking', Ouranoupolis, Greece,
May 1997; to be published in the Proceedings.}
}
\author{Jean-Jacques Dugne
\\  Laboratoire de Physique Corpusculaire \\
Universit\'{e} Blaise Pascal de Clermont-Ferrand II \\
FR-63177  Aubi\`{e}re, France
 \and Sverker Fredriksson, Johan Hansson
\\ Department of Physics\\ Lule\aa \ University of
Technology \\ SE-97187
Lule\aa , Sweden
\and Enrico Predazzi
\\ Department of Theoretical Physics \\
Universit\'{a} di Torino \\
IT-10125 Torino, Italy}

\date{ }

\maketitle

\begin{abstract}

The Higgs mechanism is the favourite cure
for the main problem with electroweak
unification, namely how to reconcile a gauge
theory with the need for massive gauge bosons.
This problem does not exist in preon models for
quark and lepton substructure with composite
$Z^0$ and $W$s, which, consequently, also avoid
all other theoretical complications and paradoxes
with the Higgs mechanism. We present a new,
minimal preon model, which explains the
family structure, and predicts several new,
heavy quarks, leptons and vector bosons.
Our preons obey a phenomenological
supersymmetry, but without so-called squarks
and sleptons, since this SUSY is effective only
on the composite scale.
\end{abstract}

\section{Introduction: Why Higgs pain?}

The Higgs mechanism is the hitherto smartest,
and maybe the only logically consistent
construction that circumvents the serious
problems caused by the massive weak gauge bosons.

This "medicine" has by now become a research field
of its own, but the extended theoretical efforts
have mostly revealed several new and unwanted
secondary effects, while no experimental signals
whatsoever have been found. This has not prevented
Higgs workers from assuming that the Higgs is
responsible for {\it all} fundamental masses,
through phenomenological Yukawa couplings.

Among the many Higgs pains one should mention that

\begin{itemize}
\item[-]  one single Higgs does not seem enough.
In some models there are many different Higgses.

\item[-]  it is currently fashionable to assume even
that the Higgs is composite, but it is not clear why
a particle that is not even fundamental should be
so important for the creation of mass in the Universe.

\item[-]  the "mass-from-the-Higgs" mechanism
raises more questions than it answers.
For instance: What gives the Higgs its mass?
How can the Higgs keep track of all different
fundamental masses in the particle zoo?
Therefore, one cannot hope to
reduce the number of {\it ad hoc} parameters
in particle physics with the help of a Higgs.

\item[-]  there are already some very complicated
medicines to cure those Higgs pains, the most
celebrated one being supersymmetry (SUSY).
Attractive as it might seem as a symmetry on its own,
relating fermions and bosons,
its main contribution to the Higgs mechanism
is to help cancel some divergent processes.

\item[-]  joining SUSY with the Higgs gives a "higgsino",
which should exist in the real world only mixed with
the photino and the zino (the SUSY partners of the
photon and the $Z^0$) inside the so-called neutralino,
which, in turn, is supposed to constitute the bulk of
the cosmic dark matter. One might ask why the
Universe has chosen such a complicated way to
hide its true nature, while not helping us with one
single piece of experimental evidence.

\end{itemize}

In this connection, one should not forget that the
weak gauge bosons $Z^0$ and $W^{+/-}$ have
{\it three} (partly interrelated)
ugly features compared to all other
fundamental interactions, namely that they are
very {\it massive}, highly {\it unstable} and
contain {\it electrically charged} members.
Only the former is logically explained by Higgs,
while the other two are facts of life,
and put in by hand in the electroweak theory.

\section{Why take a preon?}

In connection to the arguments given above, the most
important virtue of preons (subquarks) is that
they leave room for the $Z^0$ and $W^{+/-}$ to
be {\it composite}. This was pointed out in most
pioneering preon publications, although the actual
composition varies from model to model
(the bosons being composed of 2-6 preons).

The main arguments for a common substructure
of quarks and leptons in terms
of preons can therefore be listed:

\begin{itemize}

\item[-]  there are by now too many quarks and leptons
to leave any particle physicist comfortable.

\item[-]  there are regularities between quarks and
leptons, suggesting a common origin (but still so
many differences that preon modelling is a
nontrivial challenge).

\item[-]  most quarks and leptons are {\it unstable},
which in our view disqualifies them as fundamental
particles, although this logical problem is rarely
addressed in preon-free models for quarks, leptons and
the three-family structure.

\item[-]  preons can be used for building the weak
vector bosons, and therefore predict that the weak
force is not fundamental. It is just a "van
der Waals" leakage of some multi-preon states,
in almost exact analogy to the nuclear force being
an exchange of mostly spin-1 mesons, {\it i.e.},
quark-antiquark states. The celebrated relation
between the $Z^0$ mass and the Weinberg angle has
to do with preon couplings and masses, and not with
electroweak unification \cite{preons}.

\end{itemize}

\section{Preon models in general}

The existing preon models are often minimal in some
respect:

-	there can be a minimal number of different
preons, namely two. The prime example is the so-called
rishon model by Harari and others
\cite{harari,shupe}, with two spin-$1/2$ rishons
($T$ and $V$), but with three rishons in each (light)
quark and lepton. The vector bosons (and
even the photon) are composite six-preon states.

-	there can be a minimal number of preons
inside a quark or lepton, namely two.
Obviously, there must be preons with both
spin $1/2$ and $0$ in such models.
The prime example here is the so-called
haplon model by Fritzsch and Mandelbaum
\cite{fritzsch}. Typically, one now uses four
different preons (called $\alpha$, $\beta$,
$x$ and $y$ in \cite{fritzsch})
to build the light family of quarks and leptons.

There are some very clear problems with all
such preon models. First, they do not explain
why quarks and leptons seemingly belong to
three families. The two heavier
families are normally assumed to be
some kind of "excitations" of the light one.
Here one can think of internal radial
excitations, or an extra preon-antipreon
pair added to a light quark or lepton.
There are even suggestions that completely
different preon states exist inside one
and the same family ({\it e.g.}, that the $u$ and $d$
quarks have unequal preon numbers, or
that quarks and leptons are completely different).
In all such models it is hard to understand why
three different lepton numbers are conserved
(while quark numbers are not).

Secondly, the (almost?) massless neutrinos are not
well understood, especially not in the heavier families.
It is a paradox that preons in (maybe) the TeV mass
range can form light neutrinos. No preon model
(including ours) has been able to escape this problem.

In the literature there are several models
where the authors try to join supersymmetry
with preons. This is quite natural if there is
supersymmetry in Nature, since any
fundamental symmetry must have its
origin on the truly fundamental particle level.

Since SUSY is essentially a symmetry between
spin-$1/2$ and spin-$0$ particles it is most
conveniently introduced in models of the haplon type.
A first observation is then that the truly supersymmetric
partner of a quark or lepton, consisting of one preon
of each kind, is again a quark and a lepton.
Hence, a SUSY preon model needs no "squarks"
and "sleptons". SUSY is instead fully
implemented by "preons and spreons", which are not
observable on their own. One can naturally think
of "pseudo-SUSY partners", where only one preon is
substituted by a "spreon", but there is no reason
why such states must exist. Exceptions are the
heavy vector bosons, which in the haplon model
would turn into normal leptons, if one of their
spin-$1/2$ preons is substituted by a spin-$0$ spreon.
Hence, the "zino" is a neutrino, and the "winos"
are electrons and positrons!

It is quite impressive that a carefully selected
preon model can eliminate the need of both the
Higgs and all suggested SUSY particles (except the
photino and the gluino).

The current literature on preon SUSY seems
to originate from a paper by Pati and Salam
\cite{pati}, where it is suggested that all matter is
built from two exactly supersymmetric, massless
fields. Since these cannot be used for directly
building quarks and leptons, one has to assume that
they are "pre-preons". Several Japanese authors have
tried to build more conventional preons (with broken
or no SUSY) out of these two fundamental fields
(see \cite{matsu} for a recent example), and the result
is very similar to the haplon model \cite{fritzsch}.
As to the best of our knowledge, no efforts have
been made to implement SUSY (broken or not)
directly on the preon level.

\section{Our preons}

The essence of our new preon model is that we
introduce "extra" preons in order to construct all
three families of quarks and leptons, while still
maintaining as much symmetry and simplicity as
possible in the full preon scheme.

We start from the haplon idea of Fritzsch and
Mandelbaum \cite{fritzsch} that preons are
spin-$0$ and spin-$1/2$ objects. However,
it is crucial for our model that the lightest
family can, in fact, be built by {\it three}
preons, {\it i.e.}, one preon less than
in the haplon model. We also guess
that there is some pairwise supersymmetric
relation between preons, {\it i.e.},
as far as charges and the construction of
composite states are concerned. This preon
SUSY must, however, be broken in
the sense that the preon
masses cannot be pairwise identical.

We then use as many preon pairs as needed to build
the full three quark/lepton families, and end up with
{\it three} spin-$0$ "spreons" and
{\it three} spin-$1/2$ preons:

\vspace{3mm}
\( \begin{array}{cccc}
Charge & +e/3 & -2e/3 & +e/3 \\
S = 0 & x & y & z \\
S = 1/2 & \alpha & \beta & \delta
\end{array} \)
\vspace{3mm}

This is not a unique solution, but one of two
alternatives for constructing the lightest quark/lepton
family with the three preons $\alpha$, $\beta$ and $x$.
The other option has charges $-e/6$, $+5e/6$, $-e/6$,
which we, however, consider less attractive than the
ones chosen. Observe that our preons $\alpha$, $\beta$,
$x$ and $y$ are not identical to those in \cite{fritzsch},
although we use the same symbols. We assume that all
preons carry QCD colour, {\it i.e.}, we do not
adopt the idea in \cite{fritzsch} that only one
preon has colour, and therefore exists only in quarks.

There are several arguments, to be discussed later
in some detail, that lead to the conclusion that
either $z$ or $\delta$ is "superheavy", while the other
is just "heavy". A simple one is
that only {\it five} preons are needed to explain six
quarks/leptons. So, another option would be to use only
five preons, at the expense of the simple symmetry
in the scheme.

Before constructing the quarks and leptons we note
that the spin-$0$ preons could themselves be "di-preons"
of two spin-$1/2$ preons. The simplest (most symmetric)
scheme, which also gives correct QCD colour
charges, would then be:

\vspace{3mm}

$x = (\overline{\beta} \overline{\delta})$,
$y = (\overline{\alpha}\overline{\delta})$ and
$z = (\overline{\alpha}\overline{\beta})$.

\vspace{1mm}

If so, our model would contain only spinor preons,
and would therefore be closer in spirit to  the "rishon"
model \cite{harari,shupe}. Supersymmetry
then loses its meaning as a fundamental concept,
but could still be of phenomenological relevance,
like the "supersymmetry" between quarks and
diquarks, quoted by diquark experts (see \cite{anselm}
for a review on diquarks). The tighter the di-preon
binding, the more relevant is the "supersymmetry".

Strangely enough, the presence of a heavy $\delta$
gives {\it light} di-preons, while the light
$\overline{\alpha}$ and $\overline{\beta}$ build
up the heavy $z$. This resembles the idea promoted
by some diquark workers, that diquark formation
is relevant only when involving at least one heavy
quark. If the $(\alpha \beta)$ would, accordingly,
be {\it unbound} we would again have only five
objects ($\alpha$, $\beta$, $\delta$, $x$ and $y$),
{\it i.e.}, the minimal number required for three
families.

In the following, we will, for simplicity, keep the
symbols $x$, $y$ and $z$ for the di-preons, until their
internal composition needs to be discussed.

\section{Building leptons, quarks and vector bosons}

The preons fall into one triplet, $\bf 3_{1/2}$,
of spin $1/2$ and another one, $\bf 3_{0}$, of
spin $0$. Keeping in mind that quarks carry QCD
colour, while leptons and vector bosons are
colour-neutral, we get:

\vspace{3mm}

leptons = $\bf 3_{1/2} \times \bf 3_{0}^*$
(or $\bf 3_{1/2}^* \times \bf 3_{0}$)

quarks = $\bf 3_{1/2} \times \bf 3_{0}$

vector bosons = $\bf 3_{1/2} \times \bf 3_{1/2}^*$,

\vspace{1mm}
{\flushleft where} $^*$ signifies a triplet of
antipreons. Hence, all three species will occur in
{\it nonets}, with a possible split-up into one octet
and one singlet, according to the $SU(3)$
multiplication rule for two different triplets:
$\bf 3_{a} \times \bf 3_{b} = \bf 8_{ab} +
1_{ab}$ (if $a = b$, $\bf 3 \times 3 = 6 + 3^*$).

The similarity with the Eightfold Way for
constructing hadrons with the lightest triplet of quarks
($u$, $d$ and $s$) should not be exaggerated.
Our $SU(3)_{preon}$ symmetry is much more
broken than the $SU(3)_{uds}$, due to widely
different preon masses. And not even the
$SU(3)_{uds}$ is very helpful for understanding all
light hadron wave functions, in particular not the
singlets. So we will not copy the theoretical $SU(3)$
eigenfunctions for quarks and leptons.

Nevertheless, the prediction of nonets is
straightforward, and means that we predict the
existence of several new particles, namely {\it three}
new leptons, {\it three} new quarks and {\it six} new
vector bosons - all presumably too massive to be
seen by current experiments.

The nonets do {\it not} contain an underlying
three-family substructure. Although a certain
pattern can be seen, it has important deviations from
the conventional one. The celebrated "three families"
are therefore merely the twelve lightest quarks
and leptons that have been discovered so far.

\section{The leptons}

There are several, equally consistent, ways to
build nine leptons out of three preon-spreon pairs.
It seems to be mostly a matter of taste to
identify a superheavy one ($\delta$ or $z$).
We prefer to have a heavy $\delta$, because this
will give a more consistent description
of the three known charged leptons
and their neutrinos. Each lepton pair then
shares a common spreon, but different preons,
which means that the decay of a charged
lepton into its neutrino goes through an
exchange of a preon-antipreon state,
{\it i.e.}, a composite $W$. A superheavy
$z$ would, indirectly, require the $\tau$,
but not the $\mu$, to decay through
a break-up of a composite spreon,
which would violate the well-known
lepton universality.

It remains a certain ambiguity for the
choice of the preon content of
$(\mu,\nu_{\mu})$ versus
$(\tau,\nu_{\tau})$. However, choosing
a preon mass ordering where $\delta$ is
the heaviest preon and $z$ the heaviest
spreon, we pinpoint the $\tau$ and the
$\nu_{\tau}$ as the ones containing
the $z$. This hints at a mass relation like
$m_{\alpha},m_{\beta},m_x < m_y < m_z \ll m_{\delta}$
(although no ordering is fully consistent).

The lepton scheme now reads:
\vspace{3mm}

\( \begin{array}{lll}
\alpha \overline{x} = \nu_e & \alpha \overline{y}
= \mu^+ & \alpha \overline{z} = \nu_{\tau}\\
\beta \overline{x} = e^- & \beta \overline{y}
= \overline{\nu}_{\mu} & \beta \overline{z}
= \tau^-\\
\delta \overline{x} = \overline{\nu}_{\kappa 1} & \delta \overline{y}
= \kappa^+ & \delta \overline{z} = \overline{\nu}_{\kappa 2},
\end{array} \)

\vspace{1mm}

{\flushleft where} the bottom line
contains the three new, superheavy leptons,
two of which are somewhat arbitrarily
classified as {\it anti}neutrinos.

A number of observations can be made:

-	all known lepton decays can be understood as
reshuffling of preons into less massive states.
The conservation of three lepton numbers
$(L_e,L_{\mu},L_{\tau})$ among
those decays is equivalent to {\it preon
stability}. These three lepton numbers are
{\it not} conserved in general.

-	the decay of the hypothetical $\kappa$ does
{\it not} conserve all lepton numbers, one example
being $\kappa^+ \rightarrow \overline{\nu}_{\kappa 1}
+ \mu^{+} + \overline{\nu}_e$.

-	the scheme is not fully consistent, since the
mass-ordering between some charged leptons and
neutrinos does not follow the simple mass-ordering
of preons as given above. It seems as if the low
masses of neutrinos have to do both with their
electric neutrality and with the preon masses.

-	the non-observation of the three new leptons
means that the superheavy $\delta$ gives them a
mass in the $50$~GeV range or heavier. The two
new neutrinos must naturally be heavier than half the
$Z^0$ mass. {\it If} the $z$ and $x$ preons
are stable, these two
neutrinos should also be stable. Then they
should mimic the hypothetical, superheavy
neutralinos, {\it e.g.}, in cosmic-ray detectors.

-	consequently, the two massive neutrinos
are candidates for dark matter, if they were
produced in significant numbers at the Big Bang,
and if they are stable. We can, however,
exclude the possibility that they were produced
on equal footing with the three light
neutrinos, since a mass of only around
$20$~eV for one (stable) neutrino species
would explain the bulk of galactic dark matter.
There are three possible excuses why
the Universe is not dominated by those
superheavy neutrinos: (i) preons of different
masses were not produced in equal numbers;
(ii) the three spin-$1/2$ preons were produced
in equal numbers, but they did not bind into equally
many di-preons ($x$, $y$ and $z$); (iii) the
superheavy neutrinos are not stable (see below).

-	if the $x$, $y$ and $z$ are indeed di-preons,
the most interesting modification of what is
said above is that the three neutrinos $\nu_e$,
$\overline{\nu}_{\mu}$ and
$\overline{\nu}_{\kappa 2}$ have {\it identical}
preon contents, differing
only in the grouping into di-preons:
$\nu_e = \alpha (\beta \delta)$,
$\overline{\nu}_{\mu} = \beta (\alpha \delta)$
and $\overline{\nu}_{\kappa 2} =
\delta (\alpha \beta)$ (no other leptons
have mutually identical preon contents).
This opens up for $\nu_e \leftrightarrow
\overline{\nu}_{\mu}$ oscillations
(if Nature can somehow make up for the
apparent helicity difference). There can also
be decays of the heavier neutrinos,
in particular the superheavy
ones, {\it e.g.},
$\overline{\nu}_{\kappa 2} \rightarrow \tau^-
+ \overline{\nu}_{\tau}+ \mu^{+}$.
Unless the two lightest neutrinos are
massless, there should even exist
{\it electromagnetic decays},
like $\overline{\nu}_{\mu} \rightarrow
\nu_e + \gamma$, which would be the preonic
equivalent of the decay
$\Sigma^0 \rightarrow \Lambda^0 + \gamma$
in the quark world.

- the model therefore provides a way to
understand the apparent lack of
atmospheric muon-neutrinos \cite{atmos},
via either decays or oscillations among
neutrinos. The recent evidence from
the Los Alamos
LSND collaboration \cite{LSND} of a
$\nu_{\mu} \rightarrow \nu_e$ transition
can, however, be understood
only if we have wrong preon
assignments for the ambiguous choice of
$(\mu,\nu_{\mu})$ versus $(\tau,\nu_{\tau})$
(see above). If so, we would instead
get a natural oscillation (or electromagnetic
decay) between $\overline{\nu}_{\tau}$ and
$\nu_e$ and a possible small-scale
oscillation between $\nu_{\mu}$
and $\nu_e$ due to a quantum-mechanical
$\alpha - \delta$ preon mixing.
The latter will be discussed also in the
next section.

\section{The quarks}

Building quarks with our preon scheme turns out to
give several completely new possibilities, but also
some new problems. The special features of quarks,
in fact, {\it require} the spin-$0$
spreons to be di-preons. We therefore give the
suggested assignments directly in terms of
three-preon states with spin-$1/2$ preons only
(noting that $\beta \delta = \overline{x}$,
$\alpha \delta = \overline{y}$ and
$\alpha \beta = \overline{z}$):

\vspace{3mm}

\( \begin{array}{lll}
\alpha (\overline{\beta} \overline{\delta}) = u &
\alpha (\overline{\alpha} \overline{\delta}) = s &
\alpha (\overline{\alpha} \overline{\beta}) = c \\
\beta (\overline{\beta} \overline{\delta}) = d &
\beta (\overline{\alpha} \overline{\delta}) = X &
\beta (\overline{\alpha} \overline{\beta}) = b \\
\delta (\overline{\beta} \overline{\delta}) = t &
\delta (\overline{\alpha} \overline{\delta}) = g &
\delta (\overline{\alpha} \overline{\beta}) = h,
\end{array} \)

\vspace{1mm}

{\flushleft $X$}, $g$, $h$ being new quarks
with charges $-4e/3$, $-e/3$, $+2e/3$,
respectively.

Several crucial observations can now be made:

-	the three quarks in the bottom line are
superheavy, due to the presence of the $\delta$.
This pinpoints the $t$ as a {\it superheavy}
quark, {\it unrelated} to the $b$, and hence
explains the enormous mass differences in the
$b-t$ "family" and in the $\tau - t$ "pair".
These mass differences are, in
fact, a problem in any composite model of
quarks and leptons that relies on the
conventional family structure.

-	we predict the existence of two new
superheavy quarks - the $g$ ("gross")
and the $h$ ("heavy") - which are supposed
to have masses not too far beyond
the top mass.

-	the predicted $X$ quark is crucial for the
model. Judging from its position in the scheme,
it should, in some respect, lie between
$s$ and $g$, and also between $d$ and $b$.
One can, naturally, exclude that its mass falls
between those of the $d$ and $b$ quarks. Taking
into account that the lepton masses do not follow
the preon scheme in a consistent way, this is at
least not an isolated paradox. Maybe,
the neutrinos are "too light" due to their
lack of charge, while the $X$ quark is
"too heavy" due to its high charge.
We expect, however, that
"$\delta-$free" quarks and leptons are
considerably lighter than the superheavy ones.
This means that the $X$ should be lighter than
the $t$. At first sight, this prediction seems
controversial, but the experimental situation is,
unfortunately, obscured by the fact that
existing searches for new quarks have been
focused on finding a "fourth family", where the
lightest member has been assumed to be
a "$b'$" quark with charge $-e/3$ \cite{rpp}.
For instance, a recent search \cite{abachi}
for the $b'$ relies on the assumption that
the decay is dominated by the flavour-changing
neutral-current process $b' \rightarrow b \gamma$,
and the trigger system is set on energetic gammas.
Our new $X$ quark would, most likely, decay
through $X \rightarrow b + \mu^- + \nu_{\tau}$
(or $X \rightarrow b + \tau^- + \nu_{\mu}$),
which means that one should look for fast
{\it muons} correlated to $b$ quarks. There
are similar decay channels to $s$ and $d$ quarks,
and also purely hadronic ones, like
$X \rightarrow s + \overline{u} + d$
but these should be harder to find against
the background.

A completely different idea is that the discovered
"top quark" is indeed our $X$, {\it i.e.}, that the
"top" has charge $-4e/3$ and not $+2e/3$ as has been
taken for granted. There seems to be no experimental
hints as to the charge of the top, due to difficulties
with estimating the net charge of a hadronic (quark)
jet. In cases where, for instance, an outgoing
positron has been detected, one does not know if
the decaying quark was a top or an anti-top, since
the correlated $b$ quark could equally well have been
an $\overline{b}$ (from the decay of an $\overline{X}$).

Finally, one can blame a non-observation of the
$X$ on the well-known problem with
"$SU(3)$ singlets" in other models. In the
original $SU(3)_{flavour}$ model of the
$u$, $d$ and $s$ quarks, the masses
and compositions of the baryon and meson
singlets are not well understood.

-	the fact that quarks, but apparently not leptons, mix
with each other according to the
Cabibbo-Kobayashi-Maskawa matrix can be
understood in our model, but probably not in any
other preon model where quarks and leptons
are simply related. The important difference between
quarks and leptons is that the former are composed
of both preons and antipreons, giving access to
preon-antipreon {\it annihilation} channels between
certain quarks. In particular, the
$s = \alpha (\overline{\alpha} \overline{\delta})$
and the
$d = \beta (\overline{\beta} \overline{\delta})$
can mix via the process
$\beta \overline{\beta} \leftrightarrow \alpha
\overline{\alpha}$, which can go through a composite
$Z^0$, containing both $\beta \overline{\beta}$
and $\alpha \overline{\alpha}$, or via any
other gauge boson (gluon, "hypergluon", photon?).
The effect is weak, due to either the high $Z^0$ mass
or a very strong di-preon binding, which suppresses
the $\beta \overline{\beta}$ (and
$\alpha \overline{\alpha}$) wave function overlaps.
There are similar quark mixing channels
between $c$ and $t$, and between $b$ and $g$.

-	the smaller matrix elements, of order a
per cent or less, in the CKM matrix cannot be
understood with preon annihilation. Instead,
a weak quantum-mechanical mixing of the
$\alpha$ and $\delta$ preons is required.
They differ only in mass, and could hence
be mixed to some degree inside both quarks
and leptons.
This will give rise to a weak quark mixing
between some quarks, which probably
depends on quark masses and on the "location"
of the relevant preons (inside or outside
the di-preons).
It would also mix the $e^-$ and $\nu_e$ with the
$\tau^-$ and $\nu_{\tau}$, or, alternatively,
with the $\mu^-$ and $\nu_{\mu}$ for the
alternative preon assignments of the heavy
leptons, as discussed above.
The latter might be consistent
with the oscillation
$\nu_{\mu} \rightarrow \nu_e$ claimed in
\cite{LSND}, although we would expect
it to have a magnitude well below the
per cent level, which is on the lower
side of the experimental error bars.

-	it is still an open question if also CP violation
can be explained by any of these quark peculiarities.
An interesting observation is that the $K^0$
and $\overline{K}^0$ have {\it identical} preon contents,
although arranged in different preon/di-preon
configurations. This is no longer true if we introduce
a small $\alpha - \delta$ mixing,
which results in a $d - b$
quark mixing inside the kaons. If this preon
mixing also involves a quantum-mechanical phase,
it should result in a CP violation in the $K^0$
system. At least, the magnitude of the
$d - b$ quark mixing is
about the same as that of the CP violation
(and, possibly, as that of the neutrino oscillation
claimed by the LSND collaboration).
The $D^0$ and $B^0$ mesons do
{\it not} have the same preon contents
as their antiparticles,
whatever that means for their
CP symmetry.

\section{The vector bosons}

Vector bosons come about as bound states of a preon
and an antipreon. The most likely configurations are:

\vspace{3mm}

\( \begin{array}{lll}
(\alpha \overline{\beta}) = W^+ &
(\alpha \overline{\alpha}) - (\beta \overline{\beta})
= Z^0 & (\beta \overline{\alpha}) = W^-,
\end{array} \)

\vspace{1mm}

{\flushleft with} two other, heavier and
orthogonal, combinations ($Z'$ and $Z''$)
of $(\alpha \overline{\alpha})$,
$(\beta \overline{\beta})$ and
$(\delta\overline{\delta})$,
being equivalent to the $\omega$ and $\phi$
mesons in the quark model. There would
also be another two neutral states with
the mixed combinations
$(\alpha \overline{\delta})$
and $(\delta \overline{\alpha})$,
as well as two charged states of
$(\beta \overline{\delta})$ and
$(\delta \overline{\beta})$, all
resembling the four $K^{\ast}$ mesons
in the quark model.
There might, of course, be mixings among
these states, especially as they are all heavy
and unstable.

In any preon model with composite vector bosons
one also expects scalar partners \cite{preons}.
The fact that they have not been observed can
be blamed either on high masses or on very weak
couplings to quarks and leptons. The latter might
seem more realistic since spin-$0$ systems are
normally lighter than those with spin $1$
(the deuteron being an important exception).
The similarity with the mesons in the quark
world, and the considerable mass
difference between the $\pi$ and the $\rho$
mesons make us suspect
that the missing scalars might be "superlight",
and maybe even lie in the MeV mass range.
It could be worthwhile to study possible
decay modes and look for evidence in
existing experimental data. One possibility is
that the normal scalar mesons are indeed
hybrids of quark-antiquark and
preon-antipreon states.

\section{Conclusions}

Our efforts to construct a minimal preon model
for all known quarks and leptons have led us
to a scheme with only three fundamental spin-$1/2$
preons, which prefer to form tightly bound,
scalar, di-preons. Quarks and leptons are three-preon
(preon/di-preon) states, giving three new,
very heavy leptons and three new quarks,
one of which has charge $-4e/3$
and could be lighter than (or identical to)
the top quark.

The different preon contents of quarks and
leptons explain the main mixings
described by the CKM matrix, and leaves
an opening for understanding CP
violation in terms of preon mixing.

The model by itself implies (requires)
the so-called standard model of the electroweak
interaction, the Higgs mechanism, the
three families, etc, to be more or less wrong.

Many problems remain to be solved. One is the
classical paradox of light neutrinos (and light
leptons in general). Another is our
not fully consistent mass-ordering of preons. There
is hence a long way to go before we understand preon
dynamics and forces, quark/lepton wave functions
and the role of normal QCD contra a hypothetical
"hyper-QCD" with "hypergluons" that keeps
quarks and leptons together. Maybe normal QCD
is nothing but the "long-range" tail of the
hyper-QCD that acts between preons, reaching
out from the coloured quarks,
but not from the leptons.

As long as preon models have not yet developed into
something as complicated as the Higgs mechanism, we
consider them worthy of continued contemplation and
theoretical efforts, especially as there exist interesting
predictions also for energy ranges that do not need
new accelerators, or a new millenium.

\section{Acknowledgements}

We acknowledge an illuminating
correspondence with H. Fritzsch. S.F. and J.H.
are grateful to Universit\'{e} Blaise Pascal for
hospitality during visits to Clermont-Ferrand. J.-J.D.,
S.F. and E.P. would like to thank Argyris Nicolaidis
for organising a most successful and inspiring
conference in Ouranoupolis.
This project is supported by the European
Commission under contract CHRX-CT94-0450,
within the network
"The Fundamental Structure of Matter".


\begin{thebibliography}{45}

\bibitem{preons} I.A. D'Souza, C.S. Kalman,
{\it Preons} (Singapore, World Scientific 1992).

\bibitem{harari} H. Harari, N. Seiberg,
Phys. Lett. B {\bf 98}, 269 (1981);
Nucl. Phys. B {\bf 204}, 141 (1982).

\bibitem{shupe} M.A. Shupe,
Phys. Lett. B {\bf 86}, 87 (1979).

\bibitem{fritzsch} H. Fritzsch, G. Mandelbaum,
Phys. Lett. B {\bf 102}, 319 (1981).

\bibitem{pati} J.C. Pati, A. Salam,
Nucl. Phys. B {\bf 218}, 109 (1983).

\bibitem{matsu} T. Matsushima,
Nagoya University reports DNPU-89-50 (1989)
and TMI-97-1 (1997), unpublished.

\bibitem{anselm} M. Anselmino, E. Predazzi,
S. Ekelin, S. Fredriksson, D.B. Lichtenberg,
Rev. Mod. Phys. {\bf 65}, 1199 (1993).

\bibitem{atmos} Y. Fukuda {\it et al.},
Phys. Lett. B {\bf 335}, 237 (1994);
B {\bf 280}, 146 (1992);
B {\bf 205}, 416 (1988);
D. Casper {\it et al.},
Phys. Rev. Lett. {\bf 66}, 2561 (1991);
Phys. Rev. D {\bf 46}, 3720 (1992);
W. Allison {\it et al.},
Phys. Lett. B {\bf 391}, 491 (1997).

\bibitem{LSND} C. Athanassopoulos {\it et al.},
Phys. Rev. C {\bf 54}, 2685 (1996);
Phys. Rev. Lett. {\bf 77}, 3082 (1996);
nucl-ex/9709006.

\bibitem{rpp} R.M. Barnett {\it et al.},
{\it Review of Particle Physics},
Phys. Rev. D {\bf 54}, 1 (1996).

\bibitem{abachi} S. Abachi {\it et al.},
Phys. Rev. Lett. {\bf 78}, 3818 (1997);
and references therein for other
"fourth-family" quark searches.

\end{thebibliography}
\end{document}